\documentclass[a4paper]{jpconf}
\usepackage{graphicx}

\usepackage{pb-diagram}
\usepackage{latexsym}
\usepackage{amsfonts}
\usepackage[usenames,dvipsnames]{xcolor}
\usepackage{amssymb,amsmath,epsfig}
\usepackage{dcolumn}
\usepackage{bm}

\definecolor{myblue}{rgb}{0,0,0.8}
\usepackage[colorlinks=true
,urlcolor=myblue	
,anchorcolor=myblue
,citecolor=myblue
,filecolor=myblue
,linkcolor=myblue
,menucolor=myblue
,pagecolor=myblue
,linktocpage=true  
]{hyperref}

\def\nonu{\nonumber}
\def\br{\begin{eqnarray}}
\def\er{\end{eqnarray}}
\def\be{\begin{equation}}
\def\ee{\end{equation}}

\def\({\left(}
\def\){\right)}
\def\[{\left[}
\def\]{\right]}

\def\a{\alpha}

\def\b{\beta}

\def\cA{{\cal A}}
\def\cV{{\cal V}}
\def\d{\delta}

\def\bpsi{\bar{\psi}}

\def\vareps{\varepsilon}

\def\g{\gamma}

\def\l{\lambda}
\def\L{\Lambda}

\def\o{\over}

\def\p{\phi}
\def\P{\Phi}
\def\pa{\partial}

\def\smu{\sqrt{\mu}}

\def\th{\theta}
\def\bth{\bar{\theta}}

\def\tp0{\Theta_{+}^{(0)}}
\def\tm0{\Theta_{-}^{(0)}}

\def\bD{\bar{D}}
\def\bcA{\bar{\cal A}}
\def\bA{\bar{A}}

\def\bpa{\bar{\partial}}
\def\beps{\bar{\varepsilon}}

\begin{document}
\title{Type-II defects in the super-Liouville theory}

\author{A R Aguirre}

\address{Instituto de F\'isica Te\'orica -- IFT/UNESP, Rua Dr. Bento Teobaldo Ferraz 271, Bloco II, 01140-070, S\~ao Paulo, Brazil}

\ead{aleroagu@ift.unesp.br}

\begin{abstract}
The introduction of type-II defects is discussed under the Lagrangian formalism and Lax representation for the $N=1$ super-Liouville model. We derive a new kind of super-B\"acklund transformation for the model and show explicitly the conservation of the modified energy and momentum, as well as supercharge.
\end{abstract}

\section{Introduction}

The study of two-dimensional integrable field theories in the presence of defects or impurities has evolved into a rich subject in recent years from both classical and quantum points of view \cite{Del}-\cite{Niko}. The Lagrangian formalism was initially  introduced in \cite{Corr1} to describe integrable defect conditions through a suitable local Lagrangian density located at a fixed point. The fields on either side of the defect only interact with each other at the boundary, which characterizes the type-I defects. Several types of bosonic field theories \cite{Corr1,Corr2} allow this kind of defects preserving modified charges after including some defect contributions. Their integrability can be ensured by using the well-known inverse scattering method formalism where the defect conditions corresponding to frozen B\"acklund transformations turn to be encoded in the defect matrix \cite{Cau}.

Soon after, a generalization of the original Lagrangian description was proposed in \cite{Corr09}, by allowing additional degrees of freedom associated with the defect itself, which is now called type-II defects. This alternative framework was analysed in the cases of the sine/sinh-Gordon, massive free field, Liouville and Tzitz\'eica-Bullough-Dodd models in \cite{Corr09,Ale3}. For the supersymmetric extensions of sine-Gordon model \cite{Lean1,Lean2}, and for the massive Thirring models \cite{Ale,Ale4}, those additional degrees of freedom had already appeared naturally. An important feature of this kind of defects turns to be the fact of having characteristics resembling a pair of fused type-I defects.

On the other hand, it was also proposed in \cite{Ana12} a fully algebraic approach
involving the classical r-matrix structure, as well as a modified transition matrix to describe the Liouville-integrable defects, which allowed mainly to show the involutivity of the modified charges in several integrable models \cite{Ana12, Jean, Niko}.

In this paper we will provide a Lagrangian description for the type-II defects within the $N=1$ super-Liouville model. We will propose a generalization of the super-B\"acklund transformation for the super-Liouville equation given in \cite{Kulish}, by including a chiral superfield in the formulation. We will also derive the defect matrices for the Liouville and super-Liouville models and show explicitly the conservation of the modified momentum, energy and supercharge. We also show that the conformal invariance is guaranteed after introducing the type-II defects in the models, which implies they are indeed topological defects.

\newpage
\section{Type-II defect Liouville field theory }

In this section, we will review the type-II defects in the Liouville field theory by using the Lagrangian framework, and present also the Lax formalism.

\subsection{Lagrangian description}

Let us introduce a defect located at $x=0$, with $\p_1(x,t)$ be a Liouville field on the left side of it, $\p_2(x,t)$ be also a Liouville field on the right side of it, and $\L_0(t)$ a boundary field associated with the defect itself. Then, we start with the following Lagrangian density,
\br
 {\cal L} = \th(-x) {\cal L}_1 + \th(x) {\cal L}_2 + \d(x) {\cal L}_D, \label{e2.1}
\er
with
\br
{\cal L}_p &=&\(\pa_x\p_p\)^2 -\(\pa_t\p_p\)^2 + V_p,\qquad p=1,2,\\[0.2cm]
 {\cal L}_D&=&\(\p_2\pa_t\p_1 - \p_1\pa_t\p_2\) -\L_0\pa_t(\p_1-\p_2) +(\p_1-\p_2)\pa_t\L_0 + B_0\(\p_1,\p_2,\L_0\), \label{e2.3}
\er
where the Liouville potential is given by $V_p={\mu^2}\,e^{2\p_p}$ and the defect potential $B_0$ can be decomposed into $B_0= B_0^+(\p_+-\L_0,\p_-) + B_0^-(\p_-,\L_0)$, after introducing the variables $\p_{\pm}=\p_1\pm\p_2$ \cite{Corr09}. Here $\mu$ is a scale parameter sometimes called cosmological constant. It is not difficult to show that the modified energy $E+B_0$ is conserved, where $E$ denotes the canonical energy. On the other hand, by requesting a conserved modified momentum, the defect potentials $B_0^{\pm}$ have to satisfy the non-linear relation,
\br
 \( \frac{\pa B_0^+}{\partial\phi_-} \)\(\frac{\pa B_0^-}{\pa\L_0}\)-\( \frac{\pa B_0^+}{\pa\L_0}\)\( \frac{\pa B_0^-}{\pa\phi_-}\) &=& \(V_1-V_2\),
\er
and can be adequately  written as
\br
 B_0^+ &=& -2i\mu\b^2\,e^{(\p_+ -\L_0)} , \qquad 
 B_0^- \,=\, \frac{i\mu}{\b^2} \,e^{\L_0} \(\cosh\p_- +\kappa\),
\er
where $\kappa$ is an arbitrary parameter. Besides the respective bulk field equations, namely
\br
 \pa_x^2\p_p -\pa_t^2\p_p = \mu^2 \,e^{2\p_p}, \qquad p=1,2,
\er
we also obtain the defect conditions at $x=0$,
\br
\pa \phi_+ + 2\pa_{t}\L_0 &=& -{i\mu \o \b^2}\, e^{\L_0}\sinh \p_-,  \label{dc2.13}\\
\bpa\phi_- &=& 2i\mu\b^2 \, e^{(\p_+ -\L_0)},\\
\pa\phi_- &=&-{i\mu \o \b^2}\,e^{\L_0} \(\cosh \p_- +\kappa\),\label{dc2.15}
\er
where the following light-cone notation has been used, $z= (x-t)/2$, ${\bar z} = (x+t)/2$, \mbox{$\pa\equiv \frac{\pa}{\pa z} = \pa_x -\pa_t$}, $\bpa\equiv \frac{\pa}{\pa {\bar z}}= \pa_x + \pa_t$. If these type-II defect conditions hold for every $x\in \mathbb{R}$, we get
\br
\pa (\phi_+ -\L_0) &=&-{i\mu \o \b^2}\,e^{\L_0} \sinh \p_-, \label{tII1}\\
\bpa \L_0 &=&0,\label{tII2}\\[0.1cm]
 \bpa\phi_- &=& 2i\mu\b^2 \,e^{(\p_+ -\L_0)},\\
\pa\phi_- &=&-{i\mu \o \b^2} \,e^{\L_0} \(\cosh\p_- +\kappa\),\label{tII4}
\er
the type-II B\"acklund transformations for Liouville equation which couples an auxiliary holomorphic field  $\L_0=\L_0(z)$ to the Liouville fields. From the above equations we can also find an anti-holomorphic functional, namely,
\br
 \pa \left[e^{-(\p_+-\L_0)}\(\cosh\p_- +\kappa\)\] = 0.
\er
\noindent
Now, the modified conserved momentum is given explictly by ${\cal P} = P + \left[B_0^+ - B_0^-\right]_{x=0}$, where $P$ denotes the canonical momentum. It is worth also noting that when $\L_0=0$, the type-I defect first proposed in \cite{Corr1} is recovered, which can be described by the defect conditions,
\br
\pa \phi_+ &=& -{i\mu \o \b^2}\sinh \p_-, \qquad \quad \bpa\phi_- \,=\, 2i\mu\b^2 \,e^{\p_+}.\label{tId}
\er
In the next subsections, we will present a discussion about the conformal symmetry in the present of the defects, and the defect matrix which guarantees the existence of higher-order modified conserved quantities.


\subsection{Defect conformal symmetry}

In this subsection we discuss if the defect conditions introduced in subsection (2.1) indeed respect the conformal symmetry of the original bulk Liouville theory. The defect is called \emph{conformal} if the 
energy-momentum tensor flow is continuos across the defect, namely,
\br
 \left[T^{(1)}(z) - \overline{T}^{(1)}(\bar{z}) \]_{x=0} &=&  \left[T^{(2)}(z) - \overline{T}^{(2)}(\bar{z}) \]_{x=0}, \label{eq2.16}
\er
where $T^{(p)}(z)$ and $\overline{T}^{(p)}(\bar{z})$ for $p=1,2$, are the respective holomorphic and anti-holomorphic components of the stress tensor, and are given by \cite{Liao}
\br
 T^{(p)} = (\pa \p_p)^2 -\pa^2\p_p , \qquad \quad \overline{T}^{(p)} = (\bpa \p_p)^2 -\bpa^2\p_p .
\er
From the equations of motion, we can show that the conservation laws $\bpa T = \pa \overline{T} =0$ are properly satisfied. Now by using the type-II defect conditions, one can directly show that the holomorphic and anti-holomorphic part of the stress-tensor are continuous across the defect individually, namely
\br
 T^{(1)}\big|_{x=0} &=& T^{(2)}\big|_{x=0},\label{cd1}\\
 \overline{T}^{(1)}\big|_{x=0} &=& \overline{T}^{(2)}\big|_{x=0}.\label{cd2}
\er
Therefore the type-II integrable defect described by the conditions belong to the class of \emph{purely transmitting defects}, or sometimes called \emph{topological defects} \cite{Bac}, which is one of the extremal solutions for the gluing condition (\ref{eq2.16}). For the type-I integrable defect ($\L_0=0$), we find that only the second condition (\ref{cd2}) is still satisfied, but the first one (\ref{cd1}) differs by a total time-derivative term, namely,
\br
 T^{(1)}\big|_{x=0} &=& T^{(2)}\big|_{x=0} + \pa_t\left[2\pa \p_- +\frac{2i\mu}{\b^2}\(\cosh\p_- + \kappa\)\]_{x=0}.
\er
Then, the type-I defect is integrable but not conformal, and the $\L_0$ field can be thought of as responsible for recovering the conformal invariance of the original theory after introducing a defect. It is also worth noting that an alternative type-I defect can be considered for the Liouville theory which is derived just by exchanging $\pa \leftrightarrow \bpa$ in condition (\ref{tId}), and that satisfies the first gluing condition (\ref{cd1}) but not the second one (\ref{cd2}). Then, it is quite natural understand the type-II integrable defect for Liouville theory also as being the result of fusing (in the sense of \cite{Corr10}) these two different kinds of type-I defects, as was already suggested in the sine-Gordon case \cite{Corr09} (see also \cite{Rob} for fusing defects in the $a_r^{(1)}$ Toda models).


\subsection{Defect matrix}

Let us define the following auxiliary linear system:
\br
 \pa \Psi &=& A(\l) \Psi, \qquad  \bpa \Psi\,=\, {\bar A}(\l) \Psi,
\er
where $\Psi$ is a two-dimensional column vector, $\l$ is a spectral parameter, and $A$ and ${\bar A}$ are the Lax connections given by,
\br
 A(\l) &=&  \left[ \begin{array}{cc} -\mbox{\large$\frac{\pa \p}{2}$} & -\l\mu\,e^{\p} \\[0.2cm] 0 & \mbox{\large$\frac{\pa \p}{2}$} \end{array}
                    \right], \qquad 
 {\bar A}(\l) \,=\,  \left[\begin{array}{cc} \mbox{\large$\frac{\bpa \p}{2}$} & 0 \\[0.2cm] -\mbox{\large${\mu \o \l}$} \,e^{\p} & -\mbox{\large$\frac{\bpa \p}{2}$} \end{array}
                    \right].\label{e1.7}           
\er
Then, the Liouville field equation can be derived as a compatibility condition, or in other words from the zero curvature condition or Lax-Zhakharov-Shabat equation, namely,
\br
 \bpa A - \pa {\bar A}+ \left[A , \bar{A}\]=0.
\er
Let us now consider the defect matrix $K(\l)$ connecting two different auxiliary problems, $\Psi_1 = K(\l)\Psi_2$, by using the following ansatz,
\br
 K_{ij}(\l) =\a_{ij} +\l^{-1}\b_{ij} +\l^{-2} \g_{ij},
\er
which satisfies the differential equations,
\br
 \pa K = A_1 K - K A_2, \qquad \bpa K = {\bar A}_1 K- K {\bar A}_2.
\er
\noindent There are two kind of solutions for the $K$-matrix. The first one involves the set  $\{\a_{11},\a_{22},\b_{21},\b_{12},\g_{11},\g_{22}\}$. Then, we get
\br
 \a_{11}&=& a_{11}\,e^{\frac{\p_-}{2}},\qquad \,\,  \a_{22} =a_{11} e^{- \frac{\p_-}{2}}, \qquad\g_{11} = c_{11}\,e^{-\frac{\p_-}{2}},\qquad  \g_{22} = c_{11}\,e^{\frac{\p_-}{2}}, \qquad \,\,\, \mbox{}\\[0.1cm] 
 \b_{12} &=&  -2i\b^2{c_{11}}\,e^{(\frac{\p_+}{2} -\L_0)} , \qquad \qquad \qquad \,\b_{21}= \frac{ia_{11}}{\b^2} e^{-(\frac{\p_+}{2}-\L_0)}\(\cosh\p_- +\kappa\), 
\er
where $a_{11}$ and $c_{11}$ are two arbitrary constants. Then, the defect matrix takes the following form,
\br
 K =\left[
      \begin{array}{c c}
        a_{11} \,e^{\frac{\p_-}{2}} + \frac{c_{11}}{\l^2}  \,e^{-\frac{\p_-}{2}}& -\frac{2i\b^2}{\l}c_{11}\,e^{(\frac{\p_+}{2} -\L_0)}     			\\[0.1cm]
        \frac{ia_{11}}{\l\b^2}\,e^{-(\frac{\p_+}{2}-\L_0)}\(\cosh\p_- +\kappa\)& a_{11}\,e^{-\frac{\p_-}{2}} + \frac{c_{11}}{\l^2} e^{\frac{\p_-}{2}}
      \end{array}
  \right].
\er

The other solution involves the following set of components $\{\a_{12},\b_{11},\b_{22},\g_{21}\}$. In this case we find,
\br
 \a_{12}&=& a_{12}\,e^{(\frac{\p_+}{2}-\L_0)},\qquad  \b_{11} =\b_{22}=b_{11} \cosh\Big(\frac{\p_-}{2}\Big) , \qquad  \mbox{} \\[0.1cm]
 \g_{21} &=& c_{21}e^{-(\frac{\p_+}{2}-\L_0)}\(\cosh\p_-+\kappa\),
\er
where \,$a_{12} = -i\b^2 b_{11}$,\, $c_{21}=\frac{ib_{11}}{ 2\b^2}$. Then, the defect matrix can be written as,
\br
 K'=\left[
      \begin{array}{c c}
       \frac{b_{11}}{\l}\cosh\(\frac{\p_-}{2}\) & -i\b^2 b_{11}\,e^{(\frac{\p_+}{2} -\L_0)}  			\\[0.1cm]
        \frac{ib_{11}}{ 2\b^2 \l^2}\,e^{-(\frac{\p_+}{2}-\L_0)}\(\cosh\p_- +\kappa\)&  \frac{b_{11}}{\l}\cosh\(\frac{\p_-}{2}\)
      \end{array}
  \right].
\er
We remark that the existence of the defect matrix provides a sufficient condition to show that integrability is preserved after introducing the type-II defects described by the conditions (\ref{dc2.13})--(\ref{dc2.15}), since the defect contributions for an infinite number of modified conserved quantities can be recursively computed from the entries of the defect matrix as it was already shown explicitly in the sine-Gordon, Tzitzeica-Bullough-Dodd \cite{Ale3} and Thirring models \cite{Ale4}.


\section{Super-Liouville theory with defects}

In this section, we propose a supersymmetric extension of the type-II defect Lagrangian density described in  (\ref{e2.1})--(\ref{e2.3}) for the $N=1$ supersymmetric Liouville field theory.

\subsection{Lagrangian description}

Let us consider the supersymmetric Liouville field theory with type-II defects decribed by the following  Lagrangian density,
\br
 {\cal L} = \th(-x) {\cal L}_1 + \th(x){\cal L}_2 +\d(x) {\cal L}_D, \label{Ld3.1}
\er
with
\br
{\cal L}_p &=& \(\pa_x\p_p\)^2 - \(\pa_t\p_p\)^2 + \bpsi_p\(\pa_x-\pa_t\)\bpsi_p +\psi_p\(\pa_t+\pa_x\)\psi_p + \mu^2 e^{2\p_p} + 2i\mu e^{\p_p}\bpsi_p\psi_p ,\quad \mbox{}
\er
and  the defect Lagrangian can be written as ${\cal L}_D = {\cal L}_b +{\cal L}_f$,  with
\br 
 {\cal L}_b &=& \p_2\pa_t\p_1 - \p_1\pa_t\p_2  - \L_0\pa_t (\p_1-\p_2)+ (\p_1-\p_2)\pa_t \L_0 +  {\cal B}_0^+ + {\cal B}_0^-,\qquad \mbox{}\\[0.2cm] 
 {\cal L}_f &=& \bpsi_1\bpsi_2- \psi_1\psi_2 +(\psi_1-\psi_2)\L_1   -if_1\pa_t f_1  + {\cal B}_1^+ + {\cal B}_1^- + {\cal B}_{\L_1},
\er
where the defect potentials are given by,
\br
{\cal B}_0^+ &=& -2i\mu \b^2 e^{(\p_1+\p_2 -\L_0)}, \qquad \qquad \qquad\!  {\cal B}_0^- \,=\, {i\mu \o \b^2} \,e^{\L_0}\left[\cosh(\p_1-\p_2)+\kappa\], \\[0.1cm]
 {\cal B}_1^+  &=&  \smu \b \,e^{\frac{(\p_1+\p_2-\L_0)}{2}} (\bpsi_1+\bpsi_2) f_1, \qquad {\cal B}_1^- \,=\, {\smu \o \b}\,e^{\frac{\L_0}{2}} \cosh\Big({\p_1-\p_2\o 2}\Big)(\psi_1-\psi_2) f_1 ,\qquad \mbox{} \\[0.1cm] 
 {\cal B}_{\L_1} &=&  {\smu \o \b}\, e^{\frac{\L_0}{2}} \sinh\Big({\p_1-\p_2\o 2}\Big) \L_1f_1.\qquad \mbox{}
\er
Here, besides the bosonic field $\L_0$ the defect potentials also depend on two more auxiliary fermionic fields $\L_1$ and $f_1$. Then, the bulk field equations are
\br
 \pa\bpa \p_p &=& \mu^2 e^{2\p_p} +i\mu e^{\p}\bpsi_p\psi_p ,\\[0.1cm]
 \bpa \psi_p &=& \,\, i\mu e^{\p_p}\bpsi_p ,\\[0.1cm]
 \pa \bpsi_p &=&  \!\!-i\mu e^{\p_p}\psi_p, \qquad \quad p=1,2,
\er
and the defects conditions at $x=0$ are given as follows,
\br
\pa_x\p_1-\pa_t(\p_2 -\L_0)\!\!\! &=&\!\!\! i\mu\b^2 \,e^{(\p_1+\p_2 -\L_0)} -\frac{i\mu}{2\b^2}\,e^{\L_0}\sinh(\p_1-\p_2) - \frac{\b\smu}{4} e^{\frac{(\p_1+\p_2 -\L_0)}{2}} (\bpsi_1+\bpsi_2) f_1\nonumber \\
&&\!\!\!  - \frac{\smu}{4\b}\, e^{\frac{\L_0}{2}}\left[\sinh\Big({{\p_1-\p_2} \o 2}\Big)(\psi_1-\psi_2)+\cosh\Big({{\p_1-\p_2} \o 2}\Big)\L_1\] f_1,\qquad \mbox{}\\[0.2cm]
\pa_x\p_2-\pa_t(\p_1 -\L_0) \!\!\! &=&\!\!\!-i\mu\b^2 \,e^{(\p_1+\p_2 -\L_0)} -\frac{i\mu}{2\b^2}\,e^{\L_0}\sinh(\p_1-\p_2) + \frac{\b\smu}{4} e^{\frac{(\p_1+\p_2 -\L_0)}{2}} (\bpsi_1+\bpsi_2) f_1\nonumber \\
&&  - \frac{\smu}{4\b}\, e^{\frac{\L_0}{2}}\left[\sinh\Big({{\p_1-\p_2} \o 2}\Big)(\psi_1-\psi_2)+\cosh\Big({{\p_1-\p_2} \o 2}\Big)\L_1\] f_1,
\er
\br
\pa_t (\p_1-\p_2) \!\!\! &=&\!\!\! i\mu\b^2 \,e^{(\p_1+\p_2 -\L_0)} + \frac{i\mu}{2\b^2}\,e^{\L_0}\(\cosh(\p_1-\p_2) +\kappa \) - \frac{\b\smu}{4} e^{\frac{(\p_1+\p_2 -\L_0)}{2}} (\bpsi_1+\bpsi_2) f_1 \nonumber \\ && \!\!\! \!+ \frac{\smu}{4\b}\, e^{\frac{\L_0}{2}}\left[\sinh\Big({{\p_1-\p_2} \o 2}\Big)(\psi_1-\psi_2)+\cosh\Big({{\p_1-\p_2} \o 2}\Big)\L_1\] f_1,  \label{e3.13}\\[0.2cm]
 \psi_1 +\psi_2 &=&\L_1 + \frac{\smu}{\b}\,e^{\frac{\L_0}{2}}\cosh\Big(\frac{\p_1-\p_2}{2}\Big)f_1,\\[0.1cm]
 \psi_1-\psi_2 &=&\frac{\smu}{\b}\,e^{\frac{\L_0}{2}}\sinh\Big(\frac{\p_1-\p_2}{2}\Big)f_1,\\[0.1cm]
\bpsi_1-\bpsi_2 &=&{\smu\b}\, e^{\frac{(\p_1+\p_2 - \L_0)}{2}}\,f_1, \\[0.1cm]
\pa_t f_1 &=&   {i\smu \o 2\b} e^{\frac{\L_0}{2}}\left[\cosh\Big({\p_1-\p_2\o 2}\Big)(\psi_1-\psi_2)    +\sinh\Big({\p_1-\p_2\o 2}\Big)\L_1\] \nonumber \\ &&+{i\smu\b \o 2} e^{\frac{(\p_1+\p_2-\L_0)}{2}}(\bpsi_1+\bpsi_2).
\er
These equations agree with the type-II super-B\"acklund transformation for the super-Liouville model derived in the \ref{apB}, and can be seen has a straightforward extension of the already known B\"acklund transfomation proposed in \cite{Kulish}.  Then, these defect equations are invariant under the supersymmetry transformations,
\br
 \delta \p_p &=& \vareps\,  \psi_p +\beps\, \bpsi_p, \qquad \qquad \,\,\, \,\d \L_0  \,=\,  \vareps\,\L_1,\\[0.1cm]
 \d \psi_p    &=& -\vareps \,\pa\p_p - i\mu\, \beps\, e^{\p_p}, \qquad \,\, \d \L_1 \,=\, -\vareps \pa \L_0,\\[0.1cm]
 \d \bpsi_p  &=& - \beps \,\bpa\p_p + i\mu\, \vareps \, e^{\p_p}, \qquad \,\,\, \d f_1  \,=\, \frac{2i\vareps\smu}{\b} \, e^{\L_0 \o 2} \sinh\Big(\frac{\p_1-\p_2}{2}\Big) - 2i\smu\b\, \beps\, e^{\frac{(\p_1+\p_2 -\L_0)}{2}},\qquad \mbox{}
\er
where the parameter $\kappa$ in (\ref{e3.13}) must be set to $-1$. Now, we notice that it is possible to eliminate the Lagrange multiplier $\L_1$ and simplify the defect conditions  as follows,
\br
\pa_x \p_1 -\pa_t (\p_2 -\L_0) \!\!\! &=&\!\!\! i\mu\b^2\, e^{(\p_1+\p_2 -\L_0)} -\frac{i\mu}{2\b^2}\,e^{\L_0}\sinh(\p_1-\p_2) - \frac{\b\smu}{4} e^{\frac{(\p_1+\p_2 -\L_0)}{2}} (\bpsi_1+\bpsi_2) f_1\nonumber \\
&&  - \frac{\smu}{4\b}\, e^{\frac{\L_0}{2}}\cosh\Big({{\p_1-\p_2} \o 2}\Big)(\psi_1+\psi_2) f_1\label{stII3.28},\\[0.2cm]
\pa_x \p_2 -\pa_t (\p_1 -\L_0) \!\!\! &=&\!\!\! -i\mu\b^2\, e^{(\p_1+\p_2 -\L_0)} -\frac{i\mu}{2\b^2}\,e^{\L_0}\sinh(\p_1-\p_2) + \frac{\b\smu}{4} e^{\frac{(\p_1+\p_2 -\L_0)}{2}} (\bpsi_1+\bpsi_2) f_1\nonumber \\
&&  - \frac{\smu}{4\b}\, e^{\frac{\L_0}{2}}\cosh\Big({{\p_1-\p_2} \o 2}\Big)(\psi_1+\psi_2) f_1,
\er
\br
\pa_t(\p_1-\p_2)\!\!\! &=& \!\!\! i\mu\b^2 \,e^{(\p_1+\p_2 -\L_0)} +\frac{i\mu}{\b^2}\,e^{\L_0}\sinh\Big(\frac{\p_1-\p_2}{2}\Big) \nonumber \\ &&\!\!\!\! - \frac{\smu}{4}\left[ \b\,e^{\frac{(\p_1+\p_2 -\L_0)}{2}} (\bpsi_1+\bpsi_2)  -\frac{1}{\b}e^{\frac{\L_0}{2}} \sinh\Big({{\p_1-\p_2} \o 2}\Big)(\psi_1+\psi_2) \]\!f_1,\qquad \mbox{}\\[0.2cm]
 \psi_1-\psi_2 &=&\frac{\smu}{\b}\,e^{\frac{\L_0}{2}}\sinh\Big(\frac{\p_1-\p_2}{2}\Big)f_1,\\[0.2cm]
\bpsi_1-\bpsi_2 &=&{\smu\b}\, e^{\frac{(\p_1+\p_2 - \L_0)}{2}}\,f_1, \\[0.2cm]
\pa_t f_1 &=& {i\smu\b \o 2} e^{\frac{(\p_1+\p_2-\L_0)}{2}}(\bpsi_1+\bpsi_2) +{i\smu \o 2\b} e^{\frac{\L_0}{2}}\sinh\Big({\p_1-\p_2\o 2}\Big)(\psi_1+\psi_2) \label{stII3.33}.
\er
These relations can be derived from an alternative defect Lagrangian density ${\cal L}_D' = {\cal L}_b +{\cal L}_f'$, with
\br 
 {\cal L}'_f &=&\psi_1\psi_2  +\bpsi_1\bpsi_2 -if_1\pa_t f_1  + {\smu \o \b}\, e^{\frac{\L_0}{2}} \sinh\Big({\p_1-\p_2\o 2}\Big) (\psi_1+\psi_2) f_1 \nonumber \\&&  + \smu \b \,e^{\frac{(\p_1+\p_2-\L_0)}{2}} (\bpsi_1+\bpsi_2) f_1 .\qquad \mbox{}
\er
From these Lagrangian density we derive the corresponding canonical momentum, which is given by
\br
 P \!\!\!&=&\!\!\!\!\! \int_{-\infty}^{0}\!\!\!\! dx\Big[2(\pa_t \p_1)(\pa_x\p_1) + \bpsi_1 \pa_x\bpsi_1 - \psi_1\pa_x\psi_1\Big]  + \int_{0}^{\infty}\!\!\!\! dx\Big[2(\pa_t \p_2)(\pa_x\p_2) + \bpsi_2 \pa_x\bpsi_2 - \psi_2\pa_x\psi_2\Big].\quad\,\,\,\, \mbox{}
\er
By taking its time derivative, we get
\br
 \frac{dP}{dt} &=& \Big[\(\pa_x\p_1\)^2 + \(\pa_t\p_1\)^2 +\bpsi_1 \pa_t\bpsi_1 - \psi_1\pa_t\psi_1- \mu^2 e^{2\p_1} -2i\mu e^{\p_1} \bpsi_1\psi_1\Big]_{x=0}\nonu \\
 && \!\!\!\! - \Big[\(\pa_x\p_2\)^2 + \(\pa_t\p_2\)^2 +\bpsi_2 \pa_t\bpsi_2 - \psi_2\pa_t\psi_2- \mu^2 e^{2\p_2} -2i\mu e^{\p_2} \bpsi_2\psi_2\Big]_{x=0}. \label{dmI1}
\er
Now, by using the defect conditions (\ref{stII3.28})--(\ref{stII3.33}) the right-hand-side of (\ref{dmI1}) becomes a total time-derivative and then the modified conserved momentum is given by the combination,
\br
  {\cal P} &=& P +  (\bpsi_1 \bpsi_2-\psi_1\psi_2 ) - \left[{2i\mu \o \b^2} \,e^{\L_0}\sinh\Big(\frac{\p_1-\p_2}{2}\Big)  +2i\mu \b^2 e^{(\p_1+\p_2 -\L_0)}  \]\nonumber \\&& + \left[  {\smu \o \b}\, e^{\frac{\L_0}{2}} \sinh\Big({\p_1-\p_2\o 2}\Big) (\psi_1+\psi_2)   - \smu \b \,e^{\frac{(\p_1+\p_2-\L_0)}{2}} (\bpsi_1+\bpsi_2)\] f_1.\quad \mbox{} 
\er
Now, for the energy
\br
 E &=& \int_{-\infty}^{0} dx\Big[\(\pa_x\p_1\)^2 + \(\pa_t\p_1\)^2 +\bpsi_1 \pa_x\bpsi_1 + \psi_1\pa_x\psi_1+\mu^2 e^{2\p_1} +2i\mu e^{\p_1} \bpsi_1\psi_1\Big] \nonumber \\
 && + \int_{0}^{\infty} dx\Big[\(\pa_x\p_2\)^2 + \(\pa_t\p_2\)^2 +\bpsi_2\pa_x\bpsi_2 + \psi_2\pa_x\psi_2 +\mu^2 e^{2\p_2} +2i\mu e^{\p_2} \bpsi_2\psi_2\Big],\qquad \mbox{}
\er
the modified conserved energy is respectively given by,
\br
 {\cal E} &=&E + (\bpsi_1 \bpsi_2 + \psi_1\psi_2) + \left[{2i\mu \o \b^2} \,e^{\L_0}\sinh\Big(\frac{\p_1-\p_2}{2}\Big)-2i\mu \b^2 e^{(\p_1+\p_2 -\L_0)}  \]\nonumber \\&& + \left[  {\smu \o \b}\, e^{\frac{\L_0}{2}} \sinh\Big({\p_1-\p_2\o 2}\Big) (\psi_1+\psi_2)   + \smu \b \,e^{\frac{(\p_1+\p_2-\L_0)}{2}} (\bpsi_1+\bpsi_2)\] f_1.\quad 
\er
Besides being integrable, the defect theory is also invariant under the supersymmetry transformations,
the associated conserved supercharges are given by,
\br
 Q_{\vareps} &=& -\int_{-\infty}^{\infty} dx \(\psi \pa\p + i\mu  e^{\p}\bpsi\),  \qquad \bar{Q}_{\beps}\,=\, \int_{-\infty}^{\infty} dx \(\bpsi \bpa\p - i\mu  e^{\p}\psi\).
\er
Then, by considering now the defect conditions we found that the modified conserved supercharges are given by,
\br
 {\cal Q} &=& Q_{\vareps} -\frac{2\smu}{\b} \, e^{\L_0 \o 2} \sinh\Big(\frac{\p_1-\p_2}{2}\Big) f_1, \\[0.1cm]
  {\cal \bar{Q}} &=& {\bar Q}_{\beps} - 2\smu\b\, e^{\frac{(\p_1+\p_2 -\L_0)}{2}} f_1.
\er
Here it is worth pointing out that this is the first time that a type-II defect is encompassed within both supersymmetric and conformal model. Presumably, super-extensions of type-II integrable defects for some affine Toda models could be found, for instance for super sinh-Gordon model.


\subsection{Defect superconformal symmetry}

The defect is called \emph{superconformal} if both the stress tensor and the supercurrents, which are the  generators of the supersymmetry transformations, are continuous across the defect. These conditions are written as
\br
 \left[T^{(1)}(z) - \overline{T}^{(1)}(\bar{z}) \]_{x=0} &=&  \left[T^{(2)}(z) - \overline{T}^{(2)}(\bar{z}) \]_{x=0}, \\[0.1cm]
 \left[J^{(1)}(z) - \overline{J}^{(1)}(\bar{z}) \]_{x=0} &=&  \left[J^{(2)}(z) - \overline{J}^{(2)}(\bar{z}) \]_{x=0},
\er
where the holomorphic and antiholomorphic components of the stress tensor and supercurrents are given by \cite{Liao},
\br
 T ^{(p)}&=& (\pa \p_p)^2 -\pa^2\p_p +\psi_p \pa \psi_p, \qquad \quad \overline{T}^{(p)}\,=\, (\bpa \p_p)^2 -\bpa^2\p_p +\bpsi_p \bpa \bpsi_p,\\
 J^{(p)} &=& \psi_p \pa \p_p -\pa \psi_p, \qquad \qquad \qquad  \quad \bar{J}^{(p)} \,=\, \bpsi_p \bpa \p_p -\bpa \bpsi_p,
\er
for $p=1,2$. From the equations of motion, we can show again that the conservation laws $\bpa T = \pa \overline{T} = \bpa J=\pa \bar{J} =0$ are indeed satisfied. Now, after some quite straightforward computations, we conclude that the type-II defect introduced in the previous subsection for the $N=1$ super-Liouville model is indeed a topological defect, i.e. each holomorphic and anti-holomorphic part of the corresponding currents satisfy the following constraints,
\br
  T^{(1)}\big|_{x=0} &=& T^{(2)}\big|_{x=0}, \qquad \qquad   J^{(1)}\big|_{x=0} \,=\,  J^{(2)}\big|_{x=0}, \label{scd1}\\
 \overline{T}^{(1)}\big|_{x=0} &=& \overline{T}^{(2)}\big|_{x=0}, \qquad \qquad  \bar{J}^{(1)}\big|_{x=0} \,=\, \bar{J}^{(2)}\big|_{x=0}.\label{scd2}
\er
Therefore, we have found that the $N=1$ super-Liouville model with type-II defect introduced by the Lagrangian density (\ref{Ld3.1}) preserves  not only supersymmetry but also the conformal symmetry.


\subsection{Lax representation}

Let us first describe the super-Liouville equation (\ref{sLe}) as a compatibility condition of the following two linear systems of differential equations,
\br
 D\cV = {\cal A}\cV, \qquad \bD \cV\, =\, \bcA\cV,
\er
where $D$ and $\bD$ are the superderivatives introduced in \ref{apB}, $\cV(x,\th;\l)$ is a vector-valued superfield whose components are the bosonic superfields $\cV_1$ and $\cV_2$, and the fermionic superfield $\cV_3$, and $\l$ is the spectral parameter. The super-Lax connections $\cA$ and $\bcA$ are $3\times 3$ graded matrices that can be written in the following form, 
\br
 \cA &=& \!\!-\frac{1}{2}\(D\P\) {\bf H} +\sqrt{\l \mu} \exp\(\P \o 2\){\bf F}^+, \\ 
 \bcA &=&\,\,\frac{1}{2}\(\bD\P\) {\bf H} -i\sqrt{\frac{\mu}{\l}} \exp\(\frac{\P}{2}\){\bf F}^-,
\er
where $\P$ is  a  bosonic superfield given by 
\br 
\P = \p + i\bth \bpsi +  i \th \psi  +i \bth \th F,
\er
and $\{{\bf H}, {\bf F}^{\pm}\}$ are generators of the $osp(1,2)$ Lie superalgebra (see \ref{osp}). Then, the zero-curvature condition in the superspace, namely
\br
 \bD \cA + D \bcA - \{\bcA,\cA\} =0,
\er 
allows us to recover the super-Liouville equation (\ref{sLe}). Now, if we consider the $\th$-expansions of $\cV, \cA$ and $\bcA$ respectively, we can find directly the bosonic Lax operator for the supersymmetric Liouville model. Using (\ref{se44}), we obtain
\br
 \pa \Psi = A \Psi, \qquad \bpa \Psi = \bA \Psi,
\er
where,
\br
 A &=& \!\!-\frac{1}{2}(\pa \p){\bf H}-\!\l\mu\, e^{\p} {\bf E}^+ + i\sqrt{\l \mu} \,\psi \,e^{\p/2}{\bf F}^+, \label{e6.6}\\
 \bA &=&\,\,\, \frac{1}{2}(\bpa \p) {\bf H}- \frac{\mu}{\l}\, e^{\p} {\bf E}^-\, + \sqrt{\frac{\mu}{\l}\,} \bpsi \,e^{\p/2} {\bf F}^-. \label{e6.7}
\er
It is worth noting that if the fermionic fields vanish we recover the Lax connections for the Liouville model given in the linear problem in eq. (\ref{e1.7}).


\subsection{Defect super-matrix}

Now we are interested in deriving the defect matrix for the super-Liouville theory. Let us consider the graded matrix ${\cal K}$ connecting two different configurations, namely $\Psi _1 = {\cal K}(\l) \Psi_2$, satisfying the following equations
\br 
 \pa {\cal K} = A_1{\cal K} -{\cal K} A_2, \qquad \bpa {\cal K} = \bA_1{\cal K} -{\cal K} \bA_2,\label{sgt}
\er
where $A$ and $\bA$ are given in eqs. (\ref{e6.6}) and (\ref{e6.7}). Let us consider the following ansatz for the $\l$-expansion of ${\cal K}$,
\br
 {\cal K}_{ij} =  \a_{ij} +  {\b_{ij}}\,{\l^{-1/2}} +\l^{+1/2}\,\d_{ij} .
\er
Then, by solving the differential equations (\ref{sgt}), we get

\br
{\cal K}\!\!\! &=&\!\!\!\left[ 
{\begin{array}{cc|c}
\frac{b_{11}}{\l^{1/2}} \,e^{-{\p_-\o2}} + d_{11}  \l^{1/2}\,e^{\frac{\p_-}{2}}& -2i\b^2 b_{11}\l^{1/2}\,e^{({\p_+\o2}-\L_0) }  & -\b b_{11}\,e^{\(\p_2-\L_0\)\o 2}f_1  \\[0.2cm]
\frac{2id_{11}}{\l^{1/2}\b^2}\,e^{(\L_0-{\p_+\o2}) }\sinh^2\big({\p_-\o2}\big)  
& \frac{b_{11}}{\l^{1/2}}\,e^{\frac{\p_-}{2}}+ d_{11}\l^{1/2} \,e^{-\frac{\p_-}{2}}
& \frac{id_{11}}{\b}\,e^{\(\L_0-\p_1\)\o 2}\sinh\big({\p_-\o2}\big) f_1\\[-0.3cm]\mbox{}&\mbox{}&\mbox{}
\tabularnewline\hline \mbox{}&\mbox{}&\mbox{}\\[-0.3cm]
-\frac{i d_{11}}{\b}\,e^{\(\L_0-\p_2\)\o 2}\sinh\big({\p_-\o2}\big)  f_1
 & \b b_{11}\,e^{\(\p_1-\L_0\)\o 2}\,f_1
 &  \frac{b_{11}}{\l^{1/2}}+d_{11}\l^{1/2}
\end{array}}
 \right],\quad \mbox{}
\er
where $b_{11}$ and $d_{11}$ are two arbitrary constants. Then, this ${\cal K}$-matrix can be thought of as generating the type-II defect conditions for the $N=1$ super-Liouville. In addition, besides the defect contributions to the modified conserved energy, momentum and supercharges, explicitly contributions for higher order modified conserved quantities could be also derived following \cite{Ale4}. 
\ack 
I am very grateful to the organisers of the XXIst International Conference ISQS21 for the oportunity to present these ideas, and to J.F. Gomes and A.H. Zimerman for valuable discussions. I would like also to thank FAPESP S\~ao Paulo Research Foundation for financial support under the PD Fellowship 2012/13866-3.

\newpage
\appendix

\section{The $osp(1,2)$ Lie superalgebra}
\label{osp}

The $osp (1,2)$ Lie superalgebra contains three bosonic generators ${\bf H}, {\bf E}^{\pm}$ which correspond to the $sl(2)$ Lie algebra, and two fermionic generators ${\bf F}^{\pm}$. The three-dimensional matrix representation is given below
\br
{\bf H}=\left[ 
{\begin{array}{cc|c}
1 & 0 & 0 \\
0 & -1 &  0 \tabularnewline\hline
0  & 0 & 0
\end{array}}
 \right] , \quad
{\bf E}^{+}=\left[ 
{\begin{array}{cc|c}
0 & 1 & 0 \\
0 & 0 &  0 \tabularnewline\hline
0  & 0 & 0
\end{array}}
 \right] , \quad
{\bf E}^{-}=\left[ 
{\begin{array}{cc|c}
0 & 0 & 0 \\
1 & 0 &  0 \tabularnewline\hline
0  & 0 & 0
\end{array}}
 \right]
\er
\vskip -0.5cm
\br
{\bf F}^+=\left[ 
{\begin{array}{cc|c}
0 & 0 & 1 \\
0 & 0 &  0 \tabularnewline\hline
0  & 1 & 0
\end{array}}
 \right] 
, \quad 
{\bf F}^-=\left[ 
{\begin{array}{cc|c}
0 & 0 & 0 \\
0 & 0 & -1 \tabularnewline\hline
1  & 0 & 0
\end{array}}
 \right] \, .
\er
Usually this superalgebra is labelled $B(0,1)$ and describe the simplest example of a superconformal Toda theory based on a contragredient Lie superalgebra \cite{Kac, Olshan}. The (anti) commutation relations are given by,
\br
\left[ {\bf H}, {\bf E}^{\pm} \right] &=& \pm 2{\bf E}^{\pm}, \qquad \qquad\,
\left[{\bf H}, {\bf F}^{\pm} \right] = \pm  {\bf F}^{\pm}, \\[0.2cm]
\left[{\bf E}^+,{\bf E}^{-} \right] &=& 
\left\{ {\bf F}^+, {\bf F}^{-} \right\} = {\bf H} \\[0.2cm]
\left[{\bf E}^+, {\bf F}^{-} \right] &=& - {\bf F}^+ , \,\qquad \qquad
\left[ {\bf E}^-, {\bf F}^{+}\right] = - {\bf F}^- , \\[0.2cm]
\left\{{\bf F}^+, {\bf F}^{+} \right\} &=& 2 {\bf E}^+ , \qquad \qquad
\left\{ {\bf F}^-, {\bf F}^{-} \right\} = - 2 {\bf E}^-\, .
\er

\section{Type-II super-B\"acklund transformations}
\label{apB}

Let us consider a  bosonic superfield \, $\P = \p + i\bth \bpsi +  i \th \psi  +i \bth \th F$, and the superderivatives,
\br
 D = \frac{\pa}{\pa\th} + \th \pa, \qquad \bD = \frac{\pa}{\pa\bth} + \bth \bpa, \quad D^2=\pa, \quad \bD^2 = \bpa, \qquad D\bD = -\bD D.\label{se44}
\er
Then, the $N=1$ supersymmetric Liouville equation,
\br
 D\bD \P = -i\mu e^{\P},\label{sLe}
\er
can be derived from the action,
\br
 S \!\!\!  &=&\!\!\! \int\!\! d^2z\,d \bth  d\th \left[D\P\bD\P-2i\mu e^{\P}\]=\int\!\!d^2z \!\left[\pa\p\bpa\p +\psi\bpa\psi +\bpsi\pa\bpsi +\mu^2 e^{2\p} + 2i\mu e^\p \bpsi\psi\].\qquad \mbox{}
\er
By extending the auxiliary field $\L_0$ in (\ref{tII2}) to a chiral superfield, namely $\L = \L_0 + i\th \L_1$, satisfying $\bD \L=0$, we propose a generalization of the super-B\"acklund transformation for the super-Liouville equation \cite{Kulish}, as follows
\br
 D\(\P_+ - \L \) &=& \frac{i\sqrt{\mu}}{\b}\,\Xi \,\mbox{\large{$e^{\frac{\L}{2}}$}}\cosh\Big(\frac{\P_-}{2}\Big), \qquad \quad \,\, \,\bD \L \,=\, 0 ,\\[0.1cm]
 D \P_- &=& \frac{i\sqrt{\mu}}{\b}\,\Xi \,\mbox{\large{$e^{\frac{\L}{2}}$}}\sinh\Big(\frac{\P_-}{2}\Big), \quad \qquad \bD \P_- \,=\, i\sqrt{\mu}\b \,\Xi \exp\Big(\frac{\P_+-\L}{2}\Big),  
\er
where the fermionic superfield \,\,$\Xi =f_1 +\th b_1 +\bth b_2 +\bth\th f_2$, satisfies
\br
 D \Xi &=& -\frac{2\sqrt{\mu}}{\b}\mbox{\large{$e^{\frac{\L}{2}}$}}\sinh\Big(\frac{\P_-}{2}\Big),\qquad \quad
 \bD \Xi \,=\,  2\sqrt{\mu} \b \exp\Big(\frac{\P_+-\L}{2}\Big).
\er
In components, these equations take the following form,
\br
\pa (\p_+ -\L_0)\!\! &=&\!\! -\frac{i\mu}{\b^2}\,e^{\L_0}\sinh\p_- - \frac{\smu}{2\b}\,\mbox{\large{$e^{\frac{\L_0}{2}}$}}\sinh\Big(\frac{\p_-}{2}\Big)\psi_- f_1 -\frac{\smu}{2\b}\,\mbox{\large{$e^{\frac{\L_0}{2}}$}}\cosh\Big(\frac{\p_-}{2}\Big)\L_1 f_1,\qquad \,\,\,\mbox{}\\[0.1cm]
 \psi_+ - \L_1 &=&\frac{\smu}{\b}\,\mbox{\large{$e^{\frac{\L_0}{2}}$}}\cosh\Big(\frac{\p_-}{2}\Big)f_1,\\[0.1cm]
  F_+ &=& -\frac{\smu}{\b}\mbox{\large{$e^{\frac{\L_0}{2}}$}}\left[b_2\cosh\(\frac{\p_-}{2}\) +\frac{i}{2}\sinh\(\frac{\p_-}{2}\)\bpsi_- f_1\],\\[0.1cm]
\pa\bpsi_+ &=& \frac{\smu}{2\b}\mbox{\large{$e^{\frac{\L_0}{2}}$}}\bigg[i\sinh\(\frac{\p_-}{2}\)\(b_1\bpsi_- +\frac{i}{2}(\L_1 f_1) \bpsi_- - b_2\psi_- - F_- f_1\) \nonumber \\&&  - \cosh\(\frac{\p_-}{2}\)\(2 f_2+ i \L_1 b_2 +\frac{1}{2}\bpsi_-\psi_- f_1\) \bigg],\qquad \mbox{}\\[0.1cm]
\bpsi_- &=&{\smu\b} \mbox{\large{$e^{\frac{(\p_+ - \L_0)}{2}}$}}\,f_1,\\[0.1cm]
 F_- &=& \smu\b \mbox{\large{$e^{\frac{(\p_+ - \L_0)}{2}}$}}\left[b_1 +{i\o 2} (\psi_+-\L_1) f_1\],\nonumber \\&=& -{i\smu \o \b} \mbox{\large{$e^{\frac{ \L_0}{2}}$}}\left[b_2 \sinh\Big({\p_-\o 2}\Big) +{i\o 2}\cosh\Big({\p_-\o 2}\Big) \bpsi_- f_1\], \\[0.1cm]
\bpa \p_- &=& i\smu\b \mbox{\large{$e^{\frac{(\p_+ - \L_0)}{2}}$}}\bigg[b_2+ \frac{i}{2} \bpsi_+f_1\bigg] , \\[0.1cm]
\bpa \psi_- &=& \frac{\smu\b}{2} \mbox{\large{$e^{\frac{(\p_+ - \L_0)}{2}}$}}\left[ 2f_2 -i b_1\bpsi_+ + i b_2(\psi_+-\L_1)+ \( iF_+ +\frac{1}{2}\bpsi_+(\psi_+-\L_1)\) f_1 \], \,\,\,\qquad \mbox{} \\[0.1cm]
 b_1 &=& -{2\smu \o \b}  \mbox{\large{$e^{\frac{\L_0}{2}}$}}\sinh\Big({\p_-\o 2}\Big), \\[0.1cm]
 f_2&=& {i\smu \o \b}\mbox{\large{$e^{\frac{\L_0}{2}}$}}\cosh\Big({\p_- \o 2}\Big) \bpsi_- \,\,=\,\, i\smu\b\mbox{\large{$e^{\frac{(\p_+ - \L_0)}{2}}$}}(\psi_+-\L_1), \\[0.1cm]
\pa f_1 &=& -{i\smu \o \b}\, \mbox{\large{$e^{\frac{\L_0}{2}}$}}\bigg[\cosh\Big({\p_- \o 2}\Big)\psi_- + \sinh\Big({\p_- \o 2}\Big)\L_1\bigg] , \\[0.1cm]
\pa b_2 &=& {\smu \o \b}\, \mbox{\large{$e^{\frac{\L_0}{2}}$}} \left[i\cosh\Big({\p_- \o 2}\Big) F_- +\frac{1}{2}\sinh\Big({\p_- \o 2}\Big)\bpsi_-\psi_- +\frac{1}{2}\cosh\Big({\p_- \o 2}\Big)\bpsi_-\L_1\],\\[0.1cm]
b_2 &=& 2\smu \b\,\mbox{\large{$e^{\frac{(\p_+ - \L_0)}{2}}$}},\\[0.1cm]
\bpa f_1 &=& i\smu\b \,\mbox{\large{$e^{\frac{(\p_+ - \L_0)}{2}}$}}\,\bpsi_+,\\[0.1cm]
\bpa b_1 &=& \smu \b \,\mbox{\large{$e^{\frac{(\p_+ - \L_0)}{2}}$}} \left[iF_+ + {1\o 2}\bpsi_+(\psi_+-\L_1)\], \\[0.1cm]
 \psi_- &=& \frac{\smu}{\b}\,\mbox{\large{$e^{\frac{\L_0}{2}}$}}\sinh\Big(\frac{\p_-}{2}\Big)f_1,\\[0.1cm]
 \pa\p_- &=& {i\smu \o \b}\mbox{\large{$e^{\frac{\L_0}{2}}$}}\bigg[b_1 \sinh\Big(\frac{\p_-}{2}\Big) +\frac{i}{2}\sinh\Big(\frac{\p_-}{2}\Big)\L_1 f_1  +\frac{i}{2}\cosh\Big(\frac{\p_-}{2}\Big)\psi_- f_1 \bigg], \\[0.1cm]
 \pa\bpsi_- &=& {i\smu \o 4\b}\mbox{\large{$e^{\frac{\L_0}{2}}$}}\bigg[\cosh\Big(\frac{\p_-}{2}\Big)(\L_1 f_1) \bpsi_- +\sinh\Big(\frac{\p_-}{2}\Big)(\bpsi_-\psi_-) f_1 + 2i\cosh\Big(\frac{\p_-}{2}\Big)F_- f_1  \nonumber \\
 && -2i\cosh\Big(\frac{\p_-}{2}\Big)b_1 \bpsi_- + 2i\cosh\Big(\frac{\p_-}{2}\Big)b_2 \psi_- +2i\sinh\Big(\frac{\p_-}{2}\Big)b_2 \L_1 +4f_2 \sinh\Big(\frac{\p_-}{2}\Big)\bigg],\qquad \mbox{}
\er
where we have denoted
\br
 \p_\pm = \p_1\pm\p_2, \quad \psi_\pm=\psi_1\pm\psi_2, \quad \bpsi_\pm =\bpsi_1\pm\bpsi_2, \quad F_\pm =F_1 \pm F_2.
\er
We notice that these equations can be simplify as follows,
\br
\pa (\p_+ -\L_0) &=& -\frac{i\mu}{\b^2}\,e^{\L_0}\sinh\p_- - \frac{\smu}{2\b}\,e^{\frac{\L_0}{2}}\cosh\Big(\frac{\p_-}{2}\Big)\psi_+ f_1,\qquad \mbox{}\\[0.1cm]
\pa \p_- &=& -\frac{2i\mu}{\b^2}\,e^{\L_0}\sinh^2\({\p_-\o 2}\) - \frac{\smu}{2\b}\, e^{\frac{\L_0}{2}} \sinh\Big({{\p_-} \o 2}\Big)\psi_+ f_1,\\[0.1cm]
\bpa \p_- &=& 2i\mu\b^2 \,e^{(\p_+ -\L_0)} - \frac{\b\smu}{2} \,e^{\frac{(\p_+ -\L_0)}{2}}\,  \bpsi_+ f_1, \\[0.1cm]
 \psi_- &=&\frac{\smu}{\b}\,e^{\frac{\L_0}{2}}\sinh\Big(\frac{\p_-}{2}\Big)f_1,\\[0.1cm]
\bpsi_- &=&{\smu\b}\, e^{\frac{(\p_+ - \L_0)}{2}}\,f_1, \\[0.1cm]
\pa f_1 &=& - {i\smu \o \b}\, e^{\frac{\L_0}{2}} \sinh\Big({\p_- \o 2}\Big)\psi_+ , \\[0.1cm]
 \bpa f_1 &=& i\smu\b \,e^{\frac{(\p_+ -\L_0)}{2}}\, \bpsi_+  , \\[0.3cm]
\bpa \L_0  &=& \bpa \L_1 \,=\, 0 ,
\er
after eliminating the auxiliary fermionic field $\L_1$.



\section*{References}
\begin{thebibliography}{9}
\bibitem{Del}  Delfino G, Mussardo G and Simonetti P \textit{Nucl. Phys.} B \textbf{432} (1994) 518, {\tt hep-th/9409076}; \textit{Phys. Lett.} B \textbf{328} (1994) 123, {\tt hep-th/9403049}.

\bibitem{Sale1} Saleur H, {\it Lectures on non-perturbative fied theory and quantum impurity problems I}, {\tt cond-mat/9812110}; \\ Saleur H  {\it Lectures on non-perturbative fied theory and quantum impurity problems II},  {\tt cond-mat/0007309}.

\bibitem{Corr1} Bowcock P, Corrigan E and Zambon C \textit{Int. J. Mod. Phys.} A \textbf{19S2} (2004) 82, {\tt hep-th/0305022}; \textit{JHEP} \textbf{0401} (2004) 056, {\tt hep-th/0401020}.

\bibitem{Corr2} Corrigan E and  Zambon C, \textit{J. Phys.} A \textbf{37} (2004) L471, {\tt hep-th/0407199}; \textit{Nonlinearity} \textbf{19} (2006) 1447, {\tt nli.SI/0512038}.


\bibitem{Cau} Caudrelier V, \textit{ IJGMMP} \textbf{vol.5  No. 7} (2008) 1085-108.

\bibitem{Corr09}  Corrigan E and  Zambon C, {\it J. Phys.} A {\bf 42}  (2009) 475203,  {\tt hep-th/0908.3126}.


\bibitem{Ale3}  Aguirre A R, Araujo T R, Gomes J F and  Zimerman A H  {\it JHEP} {\bf 12} (2011) 056, {\tt nlin/1110.1589}.


\bibitem{Lean1} Gomes J F,  Ymai L H, and Zimerman A H \textit{J. Phys. A : Math. Gen.} \textbf{39} (2006) 7471, {\tt hep-th/0601014}. 

\bibitem{Lean2} Gomes J F,  Ymai L H, and Zimerman A H \textit{JHEP} \textbf{03} (2008) 001.


\bibitem{Ale} Aguirre A R, Gomes J F,  Ymai L H, and Zimerman A H  {\it Proceedings of Science}, \textbf{PoS(ISFTG)031} (2009),  {\tt arXiv:0910.2888v2 [nlin.SI]}; \\Aguirre A R, Gomes J F,  Ymai L H, and Zimerman A H {\it JHEP} {\bf 02} (2011) 017, {\tt nlin/1012.1537}. 

\bibitem{Ale4} Aguirre A R {\it J. Phys. A: Math. Theor.} {\bf 45} (2012) 205205,  {\tt arXiv:1111.5249}.

\bibitem{Ana12}  Avan J and  Doikou A {\it JHEP} {\bf 01} (2012) 040, {\tt arXiv:1110.4728 [hep-th]}.

\bibitem{Jean} Avan J and  Doikou A {\it JHEP} {\bf 11} (2012) 008,  {\tt arXiv:1205.1661}; \\ Avan J and  Doikou A 2013 {\it J. Phys.: Conf. Series} {\bf 411} 012003,  {\tt arXiv:1302.2456}.

\bibitem{Niko}  Doikou A and Karaiskos N {\it Nucl.Phys.} B {\bf 867} (2013) 872, {\tt arXiv:1207.5503}.

\bibitem{Kulish} Chaichian M and  Kulish P {\it Phys. Lett} B \textbf{78} (1978) 413.


\bibitem{Liao}  Liao H C and  Mansfield P {\it Nucl.Phys.}  B {\bf 344} (1990) 696.

\bibitem{Bac} Bachas C and Gaberdiel M R \textit{ JHEP} {\bf 0411}
(2004) 065,  {\tt hep-th/0411067}.

\bibitem{Corr10}  Corrigan E and  Zambon C {\it J. Phys. A: Math. Theor.} {\bf 43} (2010) 345201, {\tt arXiv:1006.0939}.

\bibitem{Rob} Robertson C {\it Folding defect affine Toda field theories}, {\tt arXiv:1304.3129}.

\bibitem{Kac}  Kac V G {\it Adv Math.} {\bf 30} (1978) 85.

\bibitem{Olshan} Olshanetsky M A  {\it Commun. Math. Phys.} {\bf 88} (1983) 63.


\end{thebibliography}
\end{document}